\documentclass[a4paper]{article}
\usepackage[dvips]{color,graphics}
\usepackage{bm}% bold math

\def\arpc#1#2#3{{\it Ann. Rev. Phys. Chem.} {\bf #1}, #2 (#3).}
\def\prl#1#2#3{{\it Phys. Rev. Lett.} {\bf #1}, #2 (#3).}
\def\pr#1#2#3{{\it Phys. Rev. } {\bf #1}, #2 (#3).}

\def\pra#1#2#3{{\it Phys. Rev. A} {\bf #1}, #2 (#3).}

\def\jcp#1#2#3{{\it J. Chem. Phys.} {\bf #1}, #2 (#3).}

\def\acp#1#2#3{{\it Adv. Chem. Phys.} {\bf #1}, #2 (#3).}

\def\science#1#2#3{{\it Science} {\bf #1}, #2 (#3).}
\def\nature#1#2#3{ {\it Nature} {\bf #1}, #2 (#3).}

\def\wf{\vert \nabla U\vert^2}

\def\expt#1{\langle #1\rangle}
\def\be{\begin{equation}}
\def\ee{\end{equation}}

\def\br{{\bf r}}

\begin{document}

\begin{titlepage}

\title{QUASI-SADDLES OF LIQUIDS: COMPUTATIONAL STUDY OF A BULK LENNARD-JONES
SYSTEM}

\author{{\bf Pooja Shah} and
{\bf Charusita  Chakravarty}
\thanks{Author for correspondence (E-mail:
{\tt charu@cse.iitd.ac.in)}}\\
Department of Chemistry,\\
Indian Institute of Technology-Delhi,\\
New Delhi: 110016, India.}
\date{\ }
\maketitle
\end{titlepage}

\vfill\eject

            \begin{center} {\bf Abstract} \end{center}
Quasi-saddles or inherent saddles  of the potential energy surface, $U$, of a
liquid are defined as configurations  which correspond to 
absolute minima of the pseudo-potential surface, $W =\wf$,  as
identified by a multi-dimensional minimisation procedure. 
The sensitivity of statistical properties of inherent saddles to 
the convergence criteria of 
the minimisation procedure is investigated using, as a test system,
a simple liquid bound by a quadratically shifted Lennard-Jones pair potential
with continuous zeroth, first and second derivatives at the cut-off 
distance.  The variation in statistical properties of saddles is
studied over a range of error tolerances spanning five orders of magnitude.
The largest  value of the tolerance corresponds to that used for the unshifted
LJ  liquids in a previous work (J. Chem. Phys. {\bf 115}, 8784 (2001)). 
Based on our results, it
is clear that there are no qualitative changes in statistical properties
of saddles over this range of error tolerances and even the quantitative
changes are small.  The lowest magnitude eigenvalue, $\vert \omega_0^2\vert$, 
 of the Hessian is, however, found to be very sensitive to the 
tolerance; as the tolerance  is decreased, $\vert \omega_0^2\vert$ 
is found to show 
an overall decrease. This indicates that if convergence criteria are not strict,
 absolute or low-lying minima of $W(\br )$ will be diagnosed
as having no inflexion directions.
The results also show that  it is not possible to set up an
unambiguous  numerical criterion to further classify the 
quasi-saddles into true
saddles which contain no zero curvature, non-translational normal modes
and inflexion points which have one or more zero-curvature normal mode 
directions.

\vfill\eject
\section {Introduction}

A  number of theoretical approaches have been developed to understand
the relationship between the potential energy surface (PES)  and the properties
of condensed phases. Since the potential energy surface, $U(\br )$, is
a complicated function of the $3N$ dimensional position coordinates
of an $N$-atom system, most approaches have concentrated on identifying
features of $U(\br )$ most relevant to understanding a particular phenomenon.
The most widely used of such energy landscape approaches is
inherent structure analysis which focuses on the sampling of local minima of
the PES \cite{sw82,fhs95,dw96,wdmmw}. 
To perform the inherent structure sampling, configurations
are sampled from a well-defined ensemble and are referred to
as  instantaneous configurations in this work. A steepest-descent or
equivalent local minimisation procedure is then initiated for
each instantaneous configuration in order to obtain the
nearest local minimum. All instantaneous configurations which 
quench to the same minimum or inherent structure
 are referred to as belonging to the basin of that minimum.
In general, the phase with lowest entropy and internal energy
will correspond to location of the system in the basin of the global minima
and phase transitions will correspond to occupancy of higher energy metastable 
minima. Inherent structure analysis has proved to be an extremely useful 
tool in understanding melting, the glass transition and protein folding
\cite{wdmmw,caa95,aprv,ss01,ma01,ssnss,bk97,olw97}.
It is, however,  limited by the fact that it focuses entirely on minima 
whereas an understanding of dynamical phenomena requires a knowledge of factors
which govern the passage of the system between basins of different
minima. For example, the dynamical slowing down associated with the
glass transition requires an understanding of why  rate constants
for inter-basin transitions  become essentially zero on experimental
time scales. In this context, it has recently been suggested that 
saddle configurations of the PES, which 
mark the border between adjoining basins of  minima, must play a crucial
role. Such saddle configurations will have $\nabla U =0$ and will
show negative curvature in one or more orthogonal directions. Saddles
of order 1 with just a single negative curvature direction will 
correspond to a simple transition state connecting two adjacent minima.
These stationary points of $U(\br )$ will also correspond to
absolute minima of the 
pseudo-potential surface, $W(\br )= \frac{1}{2}\vert\nabla U\vert^2$
\cite{alrss,bbczg,alpr,salrs,alrss01,gcgp}.
To identify the saddle points which play a significant role 
in the system dynamics, it has been suggested that inherent saddles,
analogous to inherent structures, can be identified
by  a steepest descent minimisation from
an instantaneous configuration to the nearest local minimum on the
pseudo-potential surface, W(\br )= $\frac{1}{2}\vert\nabla U\vert^2$, where 
$U(\br )$ is the
potential energy as a function of the 3N-dimensional position vector $\br$.
 The statistics of inherent saddles
has been recently analysed for several model glass-formers, including  a binary
Lennard-Jones mixture(BLJ), a modified monoatomic Lennard-Jones(MLJ) liquid
and  the binary soft-sphere (BSS) mixture.  Simulations show that these systems 
 undergo a transition from  saddle-dominated to
 minima-dominated dynamical regimes at the mode-coupling transition 
temperature, $T_{MC}$.
For $T> T_{MC}$, the system is localised largely in border regions
and basin-hopping is facile. Below $T_{MC}$, the average saddle order
goes to zero implying that the system occupies interiors of
the basins of local minima and  basin-hopping
becomes an activated process. 
The saddle order is found to be an essentially linear function of 
the saddle configurational   energy.
By extrapolating the linear trend one can obtain a threshold energy, $U_{thr}$,
at which the saddle order goes to zero. The average configurational
energy at the mode-coupling temperature is found to be very close to $U_{thr}$.
Since  a glass may be viewed as a very high-viscosity liquid, it is  expected 
that many of the statistical features of the stationary 
points  of the PES of glass-formers
will be shared by simple liquids. Recent studies of the stationary points of
Lennard-Jones and Morse liquids verify this conjecture \cite{sc01,sc02}.
The statistical properties of
inherent saddles are shown to carry interesting signatures steming from
variations in generic features of the interatomic potential, such 
as the range and curvature; for example, 
the  configurational energy is shown to  depend linearly
on saddle order with a slope that is proportional to the 
range of the pair potential. 
These recent studies  therefore  provide strong evidence that stationary points
of the potential energy surface constitute a physically significant
set of topographic features of the energy landscape.

The computational strategy for identifying stationary points or inherent
saddles in the above studies has been to locate the minima
of the pseudo-potential surface, $W(\br )
= \frac{1}{2}\vert\nabla U\vert^2$; within the limits of computational 
precision, the minima for which $W(\br )=0$ are taken to be stationary points
\cite{sw84}.
As pointed out in ref.\cite{dw02}, this approach must, however, be applied very
carefully because any minimisation algorithm will sample not only   
absolute minima of $W(\br )$ which correspond  to the stationary points of the 
true potential $U(\br )$  but also 
low-lying minima of $W(\br )$ 
which  correspond to inflexion points of $U(\br )$
\cite{dw02}.  At an inflexion point, the Hessian matrix, $H$, must have
one or more zero eigenvalues; the eigenvectors corresponding to these
zero eigenvalues do not correspond to basin-crossing displacements.
The question therefore arises as to whether
the physical significance of inherent saddle analysis is
compromised because, given finite numerical precision limits, it
may be difficult to distinguish between true saddles and inflexion points.
The available numerical studies of this issue are limited but they indicate
that the number of inflexion directions is always very small and does not
significantly affect the statistical estimates of saddle properties
\cite{alrss01,gcgp}.
For example, in the case of  the 256 atom MLJ system studied in ref.14,
no point with more that four inflexion directions was found.
In the liquid regime, the ratio of inflexion directions to 
negative curvature directions was  approximately 0.1. It rises to
0.8 very close to $T_{MC}$ but even then 
no major shift in the estimated value of $T_{MC}$ was seen.

Given the usefulness of stationary point analysis in understanding liquid-state
dynamics, we feel it is worthwhile to perform a detailed computational
study of a model system.
In this work, we therefore investigate  the sensitivity of
statistical properties of inherent saddles to details of
the minimisation procedure. For this purpose, we define a simple
liquid bound by a quadratically shifted Lennard-Jones pair potential
with continuous zeroth, first and second derivatives at the cut-off 
distance. This ensures that no loss of precision in the minimisation
procedure is present due to discontinuities in the pair potential at
 a finite cut-off distance. Previous
studies of stationary points of bulk liquids and glasses have 
used either unshifted pair potentials with long-range corrections 
\cite{sc01,sc02}
or quadratically shifted pair potentials with continuous zeroth and first
derivatives but discontinuous second derivatives \cite{alrss,bbczg,alpr,salrs,alrss01}.  The only exception in the literature known to us is the simulation of the binary soft-sphere system
but this may represent a special case because of the absence of any
attractive tail in the pair potential \cite{gcgp}. 
Unlike in other studies, we 
consider a one-component bulk liquid, rather than a binary mixture; these
two classes of systems differ only in the ease with which crystal nucleation 
takes place but the crucial statistical features
of stationary points are common to both glass-formers and simple liquids.
This monoatomic liquid bound by a quadratically shifted Lennard-Jones
potential  (QSLJ) defines  a test system for which we
can set reasonably demanding accuracy limits for the minimisation procedure.
The low-lying minima of the pseudo-potential, $W(\br )
= \frac{1}{2}\vert\nabla U\vert^2$, as located by the minimisation
procedure are referred to as quasi-saddles; however, unless the context requires
distinguishing clearly between true saddles and inflexion points, we also
refer to quasi-saddles as saddle configurations.
Changes in various properties of the saddle configurations are analysed
as a function of the precision limits set for the minimisation algorithm.
The possibility of setting up an unambiguous numerical criterion to
distinguish between true saddles and inflexion points is also discussed
This allows us to obtain a useful benchmark of the degree of error 
introduced by a given level of convergence in the minimisation procedure.
The results should be useful in two contexts. Firstly, it allows one to estimate
the computational effort to be expended in minimisation given 
the saddle properties of interest. Secondly, in a number of contexts, it
is of interest to study liquids which do not have a finite range e.g.
the Lennard-Jones and other well-known model systems. If
the statistical analysis of saddle configurations is to be a generally
applicable tool to study the disordered phase (liquids and glasses), it
is essential to establish to what extent such an analysis will be limited
by computational problems associated with the minimisation procedure. 
The paper is organised as follows. Section 2 contains details regarding 
the computational methods employed. Results are presented in Section 3.
Section 4 contains the conclusions.

\section{Computational Methods}

\subsection{Potential Energy Surface}
The potential energy surface (PES),
$U(\br )$, is defined as a scalar function of
the 3N-dimensional position vector, \br . The PES studied in this work is 
 pair-additive i.e.
\be
U(\br ) = \sum_{i=1}^N\sum_{j<i} u_p(r_{ij})
\ee
where $r_{ij}$ is the distance between atoms $i$ and $j$ and 
the pair potential, $u_p(r)$, is a quadratically shifted Lennard-Jones
function of the form
\begin{eqnarray}
u_p(r) &=& 4\epsilon \Biggl\{
\Bigl({\sigma\over r}\Bigr)^{12}
-\Bigl({\sigma\over r}\Bigr)^{6} \Biggr\} + ar^2 + br +c \qquad {\rm for} \
r\leq R_c\nonumber\\
&=&0 \qquad {\rm for}\ r>R_c.
\end{eqnarray}
The $a$, $b$ and $c$
parameters are determined by requiring that $u_p$, $du_p/dr$ and $d^2u_p/dr^2$
be zero at $r=R_c$. We take  $\epsilon$, $\sigma$ and
$m$ as reduced units for energy, length and mass respectively; all our 
results are given in reduced units. The pair potential defined
in equation (2) is referred to as the QSLJ potential in this paper.
The two-parameter quadratically shifted Lennard-Jones potential used
in previous studies of saddles of binary LJ mixtures
corresponds to a shifting function
$ar^2 + c$ and is referred to as the SLJ potential. The monoatomic
SLJ system was originally discussed in ref.\cite{sf73}. We fix the spherical
cut-off distance at 2.5$\sigma$ when computing the shifting parameters. Figure 1
compares the QSLJ, SLJ and unshifted Lennard-Jones  (LJ) potentials.

\subsection{NPT Ensemble Monte Carlo Simulations}

All the simulations reported in this paper were performed using the
isothermal-isobaric ensemble Monte Carlo method of McDonald \cite{fs96}.
 The partition
function of an $N$-particle system in this ensemble may be written as:
\be
Q_{NPT}={\beta P\over \Lambda^{3N}N!}\int d(\ln v) v^{N+1}
\exp (-\beta Pv) \int d{\bf s}\exp (-\beta V({\bf s};L))
\ee
where $P$ is the external pressure, $\beta=1/k_BT$ is the inverse
temperature, $\Lambda =\sqrt{h^2/2\pi mk_BT}$ is the thermal de Broglie
wavelength, $v$ and $L$ are the volume and length of the simulation cell
respectively and ${\bf s}=\br /L$ is the 3N-component position vector
scaled with respect to the length of the simulation cell. The volume, $v$,
and the particle positions, $\br$, are the Metropolis variables.
The random walk is performed using two types of moves: (i) volume
moves which sample in $\ln v$ and (ii) particle moves which displace
the spatial coordinates of a single, randomly chosen particle.
In the case of the unshifted Lennard-Jones system, long-range corrections were 
employed and the spherical cut-off distance, $R_c$, was scaled
whenever a volume move was made; the average value of $R_c\approx 2.8\sigma$. 
For the shifted potentials, the cut-off
distance was fixed at 2.5$\sigma$.

A simulation cell with 125 particles and rhombic dodecahedral boundary
conditions was used in the Monte Carlo simulations. The solid phase
structure taken to be a  face-centred
cubic (fcc) lattice.  Volume moves
constituted 7.4\% of trial moves. Acceptance ratios for volume and particle
moves were kept at approximately 50\% . Run lengths were between
two and four million configurations with equilibriation periods
of  one million to two million. This was found to be sufficient
to converge configurational averages, such as the average potential
energy, density and bond orientational parameters, to better than 1\%.
In order to obtain statistical averages over inherent saddle
configurations, 100 instantaneous configurations were sampled from the NPT-MC
simulations at equal intervals and used to generate inherent saddles. 
The simulations were carried out
at a reduced pressure of $P^*=0.67$. At this pressure, the melting point
of the unshifted Lennard-Jones system
has been established from previous work as $T^*=0.75$ \cite{hv69}. 

\subsection{Locating Inherent Saddles}

Locating inherent saddles requires one to find the absolute minima
for the pseudo-potential function, $ W(\br ) =\frac{1}{2}\wf$. Any local
minimisation technique equivalent to a steepest descent procedure
can be used to perform the minimisation. In a previous study \cite{sc01},
three minimisation procedures for locating inherent saddles were 
compared: (i) the conjugate gradient algorithm  given in ref.\cite{pftv}.
(ii) the variable metric LBFGS algorithm, as implemented  by 
Liu and Nocedal \cite{ln89} and 
(iii) Powell's direction set method which requires no
gradient information.
The LBFGS algorithm was found to be computationally the most efficient and
was therefore used  in this study  to locate local
minima in $W (\br )$ using $\nabla W = -{\bf H}\cdot {\bf F}$. 
The termination condition for the LBFGS algorithm is 
\be
\vert\vert g_i \vert\vert \leq \tau\  {\rm max}(1,\vert\vert x_i\vert\vert )
\ee
where ${\bf g}_i$ and ${\bf x}_i$ are the gradient and solution vectors
for the $i$-th iteration and $\vert\vert \cdot\vert\vert$ denotes the
Euclidean norm. Clearly the more demanding the accuracy requirements i.e.
the smaller the error tolerance $\tau$, the more computationally
demanding the minimisation procedure will be. This is illustrated in Table 1
which shows the average number of iterations required to 
converge to the nearest saddle for a given value of $\tau$ for
three different temperatures.  The error tolerance $\tau$
 is the key convergence parameter with respect to which we
wish to test the sensitivity of properties of quasi-saddles.

While absolute minima of $W(\br )$ must correspond to true stationary
points of $U(\br )$, 
local minima in $W (\br)$ need not correspond to
true stationary points of $U(\br )$ with all elements of the 
gradient vector, $\nabla
U$, equal to zero. Some local minima  may be inflexion points
where both the first and second derivatives in some directions are zero.
In principle, such flex points can be distinguished from true saddles
by looking at the absolute magnitude of $W(\br )$. However, 
given finite numerical precision it may be difficult to unequivocally
distinguish between low-lying  and absolute minima.
Alternatively, such inflexion points can be located by examining whether 
any of the eigenvalues
of the Hessian for a configuration corresponding to a minimum
in $W(\br )$, other than the three translational modes, is unusually 
small. However, the lowest magnitude, non-translational, eigenvalue
of the Hessian is very sensitive to the accuracy of the minimisation procedure.
Relaxed tolerances can lead to dramatic undercounting of the number
of flex points while rigorous convergence criteria lead to almost
99\% of local minima of $W(\br )$ being identified as inflexion points
(for example, one can contrast the results presented in refs.\cite{alrss} and
\cite{alrss01}.

\section{Results and Discussion}

\subsection{Isobars of the Quadratically Shifted Lennard-Jones liquid}

The behaviour of the qudratically shifted Lennard-Jones system (QSLJ),
as defined in Section 2.1,  is mapped out along the $P^*=0.67$ isobar
to facilitate comparison with the unshifted Lennard-Jones  system.
We have also performed NPT ensemble simulations at the same pressure
for the two-parameter quadratically shifted Lennard-Jones (SLJ) potential
with discontinuous second derivative, as defined by Stoddard and Ford\cite{sf73}.
The binary Lennard-Jones mixtures for which inherent saddle analysis
has been performed recently employ the SLJ type of cut-off conditions.
Figures 2(a) and 2(b) compare the temperature dependence of the
configurational energy per particle and the number density respectively
along the $P^*=0.67$
isobar for the QSLJ, SLJ and LJ systems. The thermodynamic melting
temperature, $T_m^*$, is known to be
0.75 for the LJ system at this pressure \cite{hv69}.
The metastability limit of the solid from our NPT simulations is
$T_k^*=0.88$; such a 15-20\% difference between $T_m^*$ and $T_k^*$ is fairly
typical. Though the difference in the well depth of the pair
potential between the LJ and SLJ systems is very small (see Figure 1),
the $T_k^*$ values for the SLJ system is about 13\% lower than
that of the unshifted LJ system because of the finite-range of the SLJ
pair potential. The QSLJ system has an even lower value of $T_k^*= 0.64$
because of the reduced well-depth. At any given temperature, the QSLJ system
has the lowest number density and highest configurational potential
energy. Since we are interested in the stationary points sampled in the
disordered liquid phase, the results of the inherent saddle analysis discussed
below refer to liquid state simulations for $T^* \geq 0.75$.

\subsection{Sensitivity of Saddle Properties to Accuracy of Minimisation
Procedure}

A saddle point is characterised by its configurational energy, $U_s$,
and its index density, $n_s$, which corresponds to the fraction of
imaginary modes; the  NPT averages are denoted by
$\expt U_s$ and $\expt n_s$.  
In the work of Broderix et al,  $\expt U_s$ and $\expt n_s$ are referred to
as parametric averages with respect to $T$ as the parameter.
For the binary LJ mixture, they showed 
 that the points $(\expt U_s, \expt n_s)$
 lie on a straight line. They also defined the geometric average, $U_s(n_s)$
as the average configurational energy of all saddles of
index density $n_s$ and showed that the 
straight line fits to the geometric and parametric averages are identical,
within the limits of statistical error.
This linear relationship between the saddle order and configurational
energy has been subsequently shown to be true for Lennard-Jones
and Morse liquids; moreover, the slope, $\partial U_s/\partial n_s$
has been found to be proportional to the range of the potential 
\cite{sc01,sc02}.
The linear form of the $\expt U_s(\expt n_s)$ curves in the disordered, liquid
phase can be understood on the basis of  a localised 
mechanism  for generating  imaginary 
frequency displacement modes which requires rearrangement of
only a few neighbouring atoms.  

Figure 3(a) shows the parameteric points $(\expt U_s, \expt n_s)$
obtained from minimisations corresponding to three different values
of the error tolerance, $\tau =10^{-10}$, $10^{-12}$ and $10^{-15}$.
It is well know that error tolerances for systems with finite
cut-offs and discontinuous seroth, first or seond derivatives
cannot be made very small. The tolerance value of $\epsilon =10^{-10}$
corresponds to that used in a previous study of the unshifted LJ
liquid \cite{sc01}. The tolerances used for the  Morse potentials
in a more recent study are similar.
The slopes of the straight line fits are almost unchanged   over 
five orders of magnitude in $\tau$. The results at $\tau =10^{-12}$
and $10^{-15}$ are  indistinguishable. For $\tau =10^{-10}$, for a 
given value of the saddle energy, the saddle order is somewhat overestimated.
Figure 3(b) shows the temperature dependence of the ensemble averaged 
index density. The behaviour with $\tau$ is similar to that seen
for the $U_s(n_s)$ curves. Figure 3(c) shows the ensemble-averaged 
saddle configurational energies as a function of temperature.
The convergence with $\tau$ is much better for the saddle energies
than for index densities. The reason is that the
convergence of the near-zero eigenvalues of the Hessian is 
relatively strongly affected by the error tolerance compared to the
overall saddle configurational energies. To substantiate this
point, we have compared the change in the absolute magnitude of the
non-translational
eigenvalues, $\vert \omega_i^2\vert$, of the Hessian with $\tau$ for 
several configurations. Figure 4 shows the values of the lowest,
$\vert \omega_0^2\vert$, and
third lowest, $\vert\omega_2^2\vert$,
 magnitude eigenvalues of the Hessian for an arbitrary sequence of
configurations for $\tau =10^{-14}$ and $10^{-15}$. The lowest magnitude
eigenvalue varies by over a factor of 10 when $\tau$ is 
reduced by an order of magnitude. There is, on average, a
decrease in the magnitude of $\omega_0^2$ as $\tau$ decreases.
The magnitudes of the
third-lowest eigenvalue (and all other higher magnitude eigenvalues)
are practically unchanged with $\tau$. The sensitivity of the lowest magnitude
eiegenvalues of the Hessian to the error tolerance of minimisation is
responsible for the relatively large error in $\expt {n}_s$ compared to
$\expt U_s$ discussed above.

\subsection{Distinguishing between True Saddles and Inflexion Points}

We have approached the possibility of setting up an unambiguous
numerical criterion to separate the quasi-saddles into 
true saddles and inflexion points by comparing the location of
each quasi-saddle on a two-dimensional plot with $x-$ and $y-$ axes
corresponding to $W (\br _s)$
and $\vert \omega_0^2\vert /\omega_E^2$ respectively
where $\omega_0^2$ is the lowest magnitude 
non-translational eigenvalue of the Hessian (see Figure 5) 
and $\omega_E$ is the Einstein frequency. 
True saddles
should correspond to very small values of $W$ and relatively
large values of $\vert\omega^2_0\vert$ and therefore should
cluster in the upper left hand corner of the graph.
Inflexion points
should correspond to relatively larger  values of $W$ and near-zero
 values of $\vert\omega^2_0\vert$ and therefore should
cluster in the lower right hand corner of the graph.
If two such well-defined clusters of points are seen, then an unambiguous
numerical distinction between true saddles and inflexion points 
can be made. Figure 5 shows, however, that the situation is quite the
opposite. Results for two different temperatures are shown and are
representative of the behaviour in other simulation runs. Only a
few points (less than 5 in 100) can be classed unambiguously as true saddlles;
the others form an almost continuous distribution.
This is presumably why in ref.16 the use of tight
convergence criteria for minimisation led to an identification of 
99\% of the quasi-saddles as inflexion points.
We suggest that the reason for this behaviour is that 
in the absence of symmetry constraints  requiring some collective  normal
mode directions of quasi-saddles
to have zero curvature, the distinction between saddles and
inflexion points is not physically very meaningful because it
attempts to distinguish between zero and near-zero curvature situations.
Coupled with the fact that the number of near-zero curvature modes is always
small, as can be seen even from the saddle normal mode spectra described
in ref.\cite{sc01} as well as from the results in ref.\cite{alrss01},
it would appear  that
the effect of reducing $\tau$ will be to introduce small
corrections in the saddle index density. Since the physical significance
of such quasi-saddles stems from the fact that the majority of
negative curvature directions will correspond to basin-crossing
displacements, improving the error tolerance will not qualitatively alter
any results.

\section{Conclusions}

The purpose of this study was to test the sensitivity of statistical properties
of quasi-saddles of the potential energy surface, $U(\br )$, to the
convergence criteria used for minimisation on the pseudo-potential surface,
$W(\br ) = \wf $. For this purpose a quadratically shifted
Lennard-Jones (QSLJ) pair potential   was defined which has continuous
zeroth, first and second derivatives at the cut-off distance. 
Inherent saddles (or quasi-saddles) of the bulk QSLJ liquid were studied as
a function of the error tolerance, $\tau$, of the LBFGS minimisation
algorithm. The variation in statistical properties of saddles was
studied over a range of $\tau$ values covering five orders of magnitude.
The largest $\tau$ values correspond to those used for unshifted
LJ and Morse liquids simulations previously. Based on our results, it
is clear that there are no qualitative changes in statistical proeprties
of saddles over this range of error tolerance and even the quantitative
changes are small.  The ensemble-averaged saddle index densities were
found to be more sensitive to $\tau$ than the corresponding averages over
saddle configurational energies. The lowest magnitude eigenvalue, $\vert \omega_0\vert^2$,  of the Hessian was, however, found to be very sensitive to the 
tolerance; as $\tau$ was decreased, $\vert \omega_0\vert^2$, was found to show 
an overall decrease. This indicates that if convergence criteria are not strict,
 absolute or low-lying minima of $W(\br )$ willl be diagnosed
as having no inflexion directions, as was found in ref.12 and 18.
For the smallest value of $\tau$, we also considered the possibility of
setting up a numerical criterion to clearly distinguish between true
saddles and inflexion points by considering the value of
both $W (\br )$ as well as $\vert \omega_0^2\vert $ at a 
quasi-saddle identified by minimisation on the pseudo-potential surface.
We find\, however, that the quasi-saddles do not fall into two such clear
and distinct categories. The reason appears to be that of the $3N$ normal mode directions associated with a low-lying or absolute minima of $W(\br )$,
a few directions will correspond to near-zero curvatures.
Deducing which of these near-zero curvatures is actually identically equal to
zero and therefore corresponds to an inflexion direction is necessary
in order to distinguish between true saddles and inflexion points
Moreover, this  appears not to be a physically very meaningful distinction
since the physically important feature of quasi-saddles is the
existence of negative curvature directions indicating their
location at the border between adjacent minima.

{\bf Acknowledgements} CC would like to thank
the Council for Scientific and
Industrial Research for financial support and the
Computer Services Centre of 
I.I.T-Delhi for access to their computational resources.
PS would like to thank the Council for Scientific and Industrial Research,
New Delhi for the award of a Senior Research Fellowship.

\vfill\eject

\vfill\eject

{\bf Table 1}

Average Number of iterations of the LBFGS minimisation algorithm 
for a given error tolerance, $\tau$, at a reduced temperature $T^*$
for the quadratically shifted Lennard-Jones (QSLJ) liquid.

\begin{tabular}{cccc}
\hline
$T^*$  &  $\tau =10^{-10}$  & $\tau = 10^{-12}$ & $\tau =10^{-15}$\\
\hline
0.67 &  66   & 1306  & 11469\\
0.88 &  54   & 1077  & 10895\\
1.29 &  35   & 898   & 17362\\
\hline
\end{tabular}

\vfill\eject

\begin{center}
{\bf Figure Captions}
\end{center}

\begin{enumerate}

\item 
Comparison of the quadratically shifted Lennard-Jones (QSLJ)
 pair potential with continuous
zeroth, first and second derivatives  with the
two-parameter quadratically shifted Lennard-Jones (SLJ) potential
defined in ref.22 as well with the unshifted Lennard-Jones (LJ) potential. 
The well-depth, $\epsilon$, and size parameter, $\sigma$, 
of the LJ potential are taken as reduced units of energy and length
respectively.

\item Dependence of configurational properties on reduced temperature $T^*$ at 
a reduced pressure $P^*$=0.67 for  the bulk LJ, SLJ and QSLJ 
systems: (a) average potential energy per particle $<U^*>$ and
(b) reduced density ${\rho}^*$

\item Statistical properties of saddle configurations for three different
values of the error tolerance, $\tau$, used in the LBFGS minimisation
algorithm: (a) Parametric averages of saddle configurational energies,
 $\expt {U}_s$, versus index densities, $\expt n_s$ (b) saddle index 
densities as a function of the reduced temperature, $T^*$ and
(c) saddle configurational energies, $\expt {U}_s$,
as a function of the reduced temperature, $T^*$. Configurational
energies $U$ are in units of $\epsilon$ per particle.

\item Sensitivity of the lowest magnitude, $\vert \omega_0^2\vert$, and
third lowest magnitude, $\vert \omega_2^2\vert$ eigenvalues of
the Hessian for a set of 20 saddle configurations sampled from
the simulation at $T^*=1.05$ for for $\tau =10^{-14}$ and $ 10^{-15}$.
The y-axis has a logarithmic scale.

\item  Scatter plots showing the correlation of the 
$\vert \omega_0^2\vert /\omega_E^2$ values with the
value of the pseudo-potential, $W(\br )= \wf$, for saddle configurations
sampled from simulations at two different temperatures where $\omega_E$ is
the Einstein frequency.

\end{enumerate}

\begin{figure}
\includegraphics{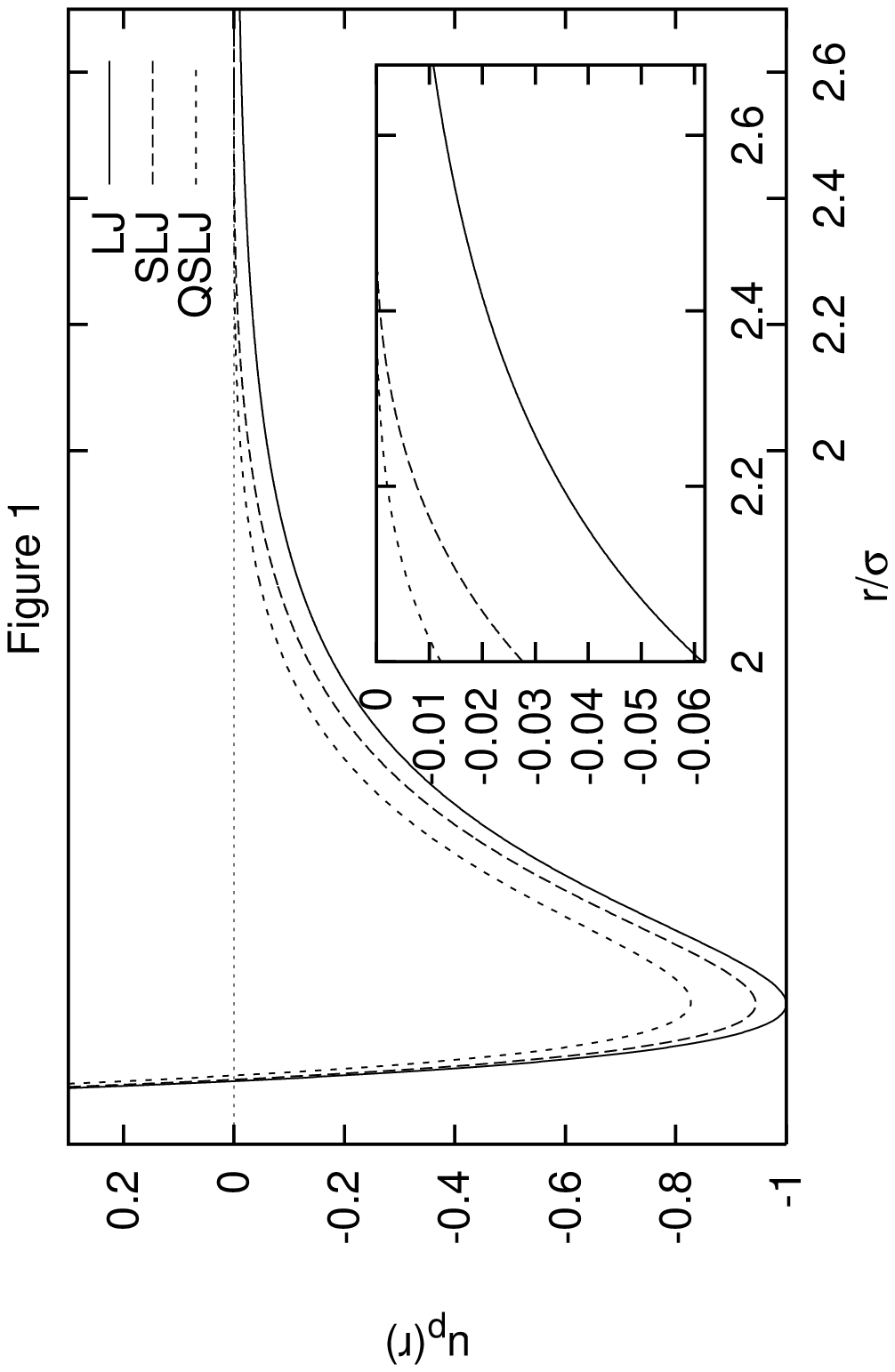}
\end{figure}

\begin{figure}
\includegraphics{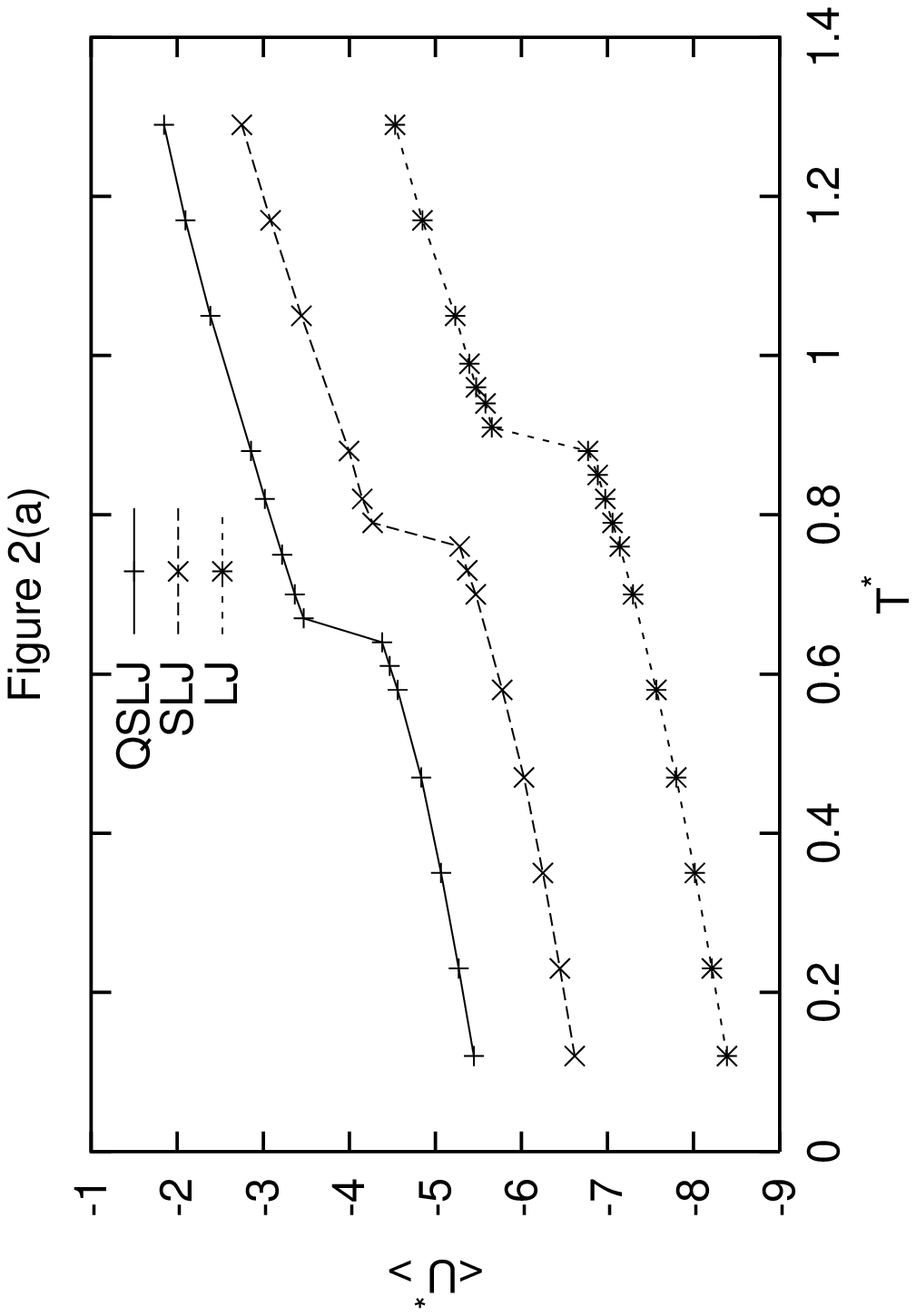}
\end{figure}

\begin{figure}
\includegraphics{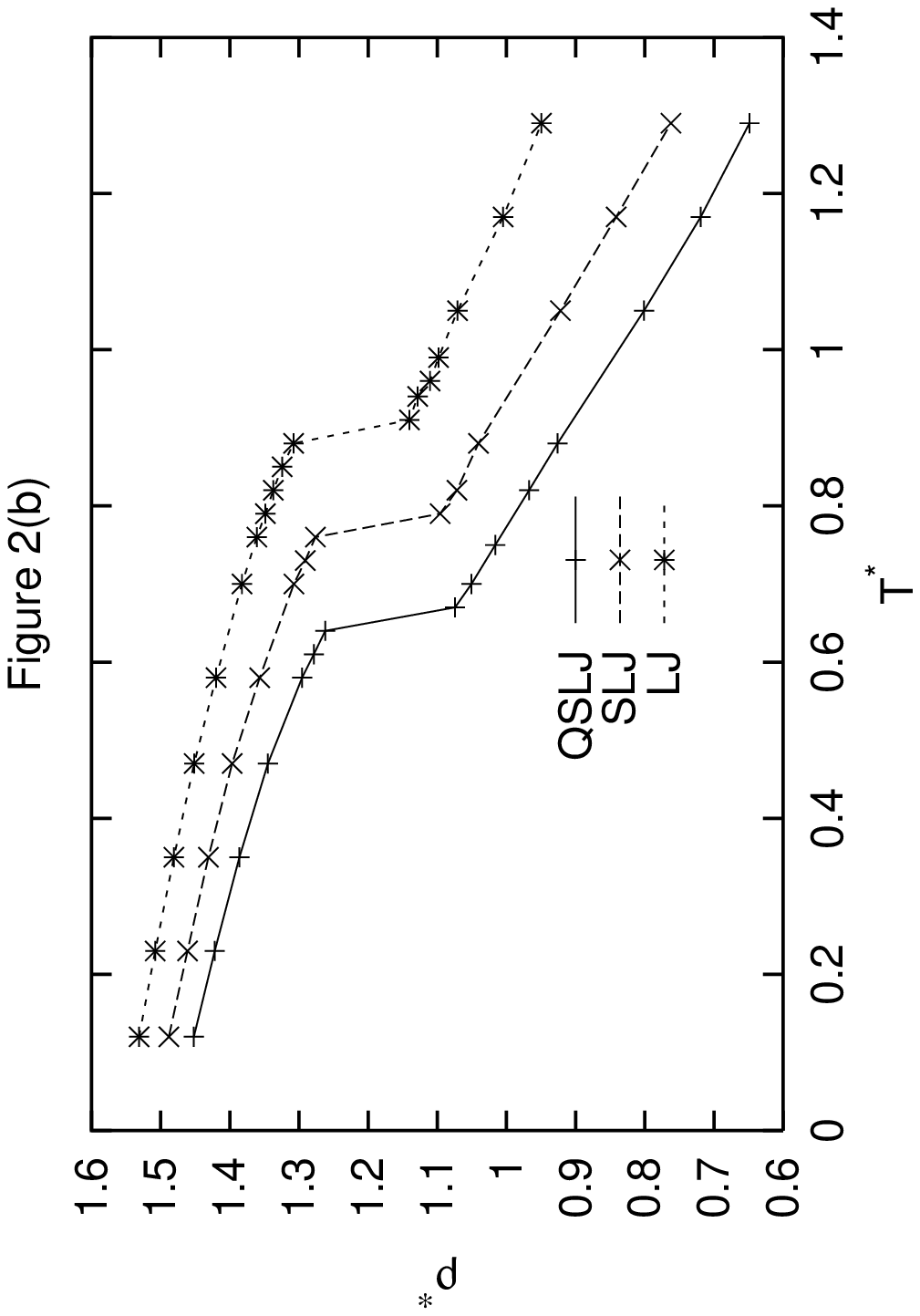}
\end{figure}

\begin{figure}
\includegraphics{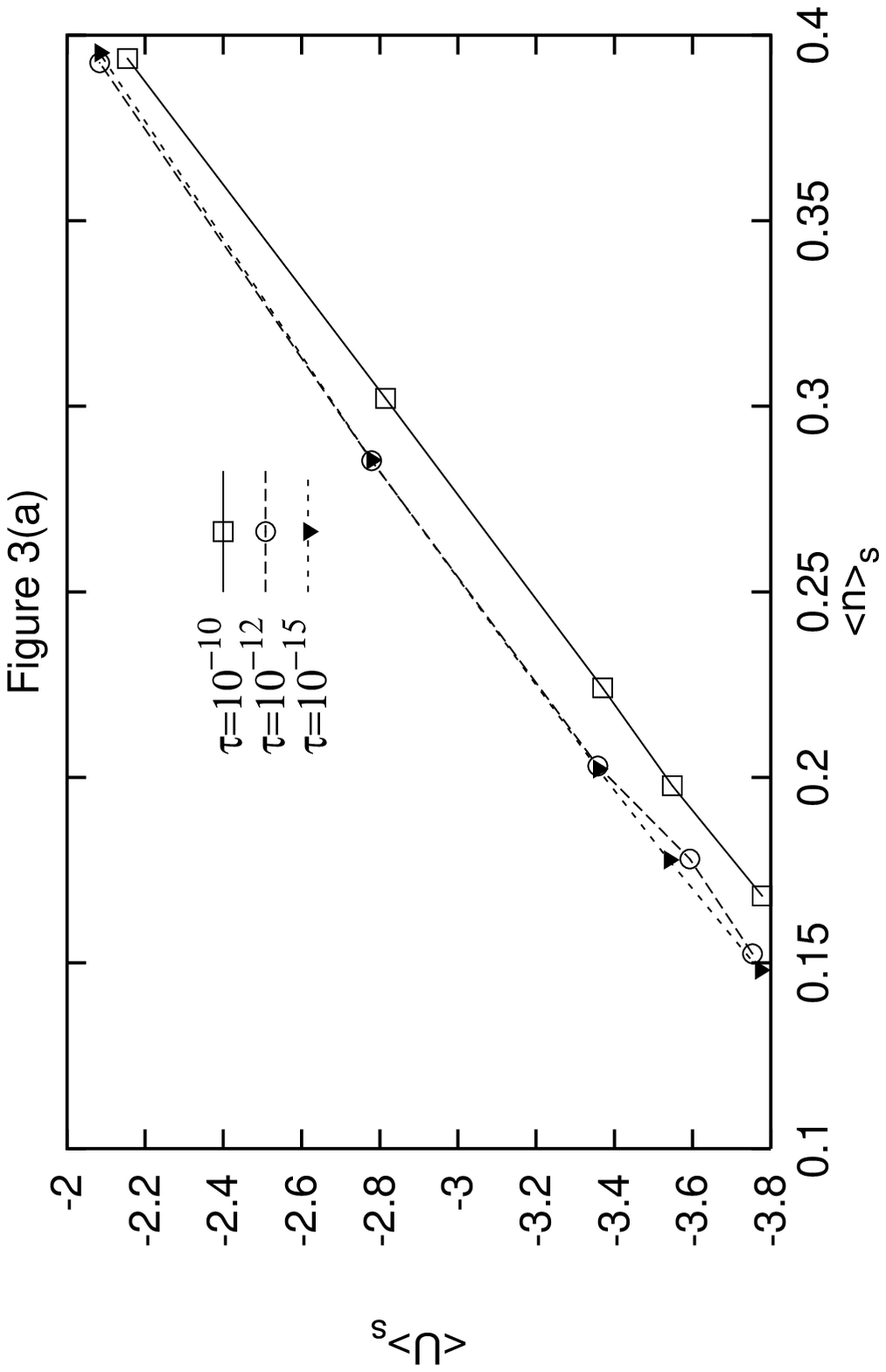}
\end{figure}

\begin{figure}
\includegraphics{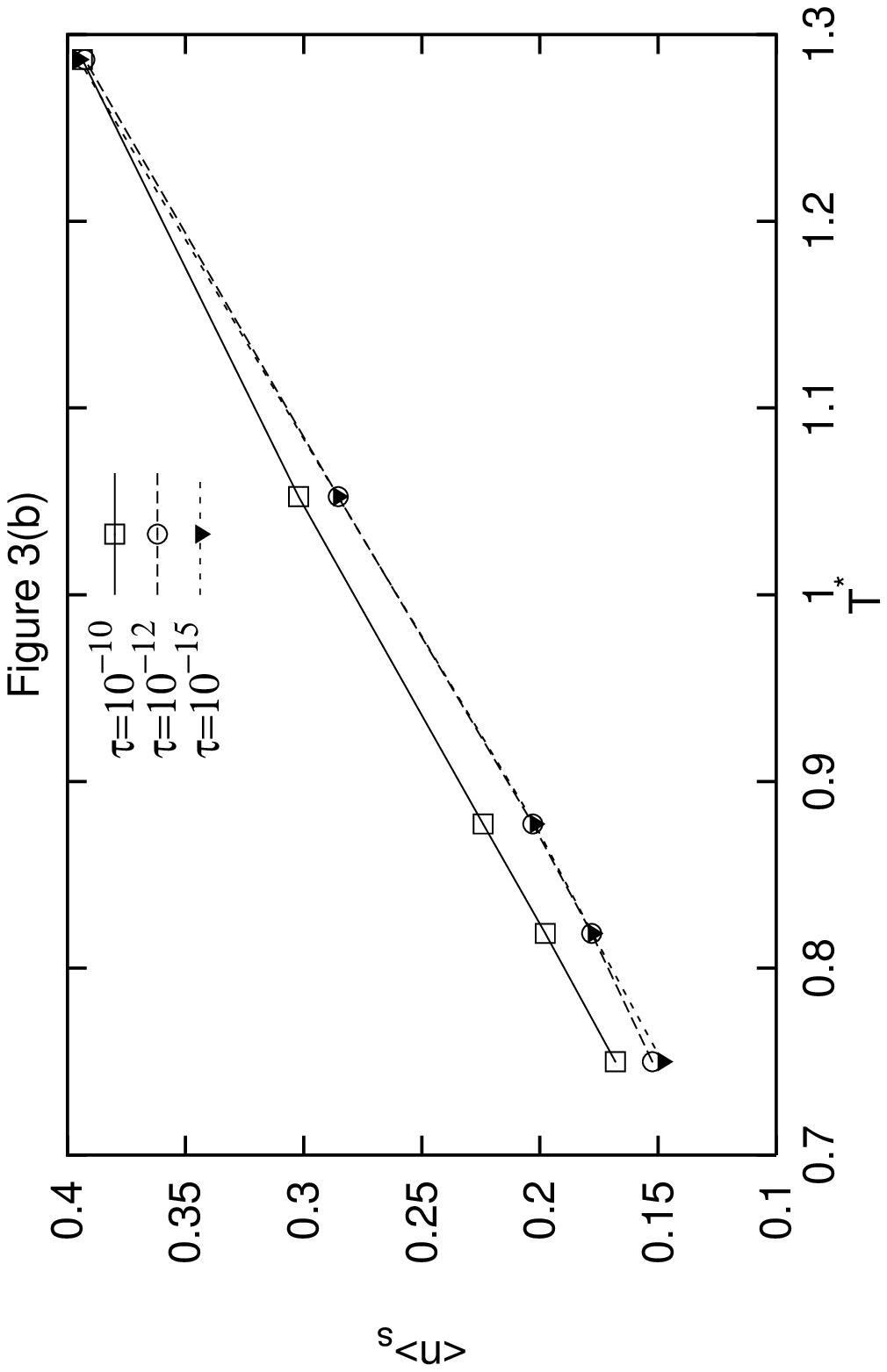}
\end{figure}

\begin{figure}
\includegraphics{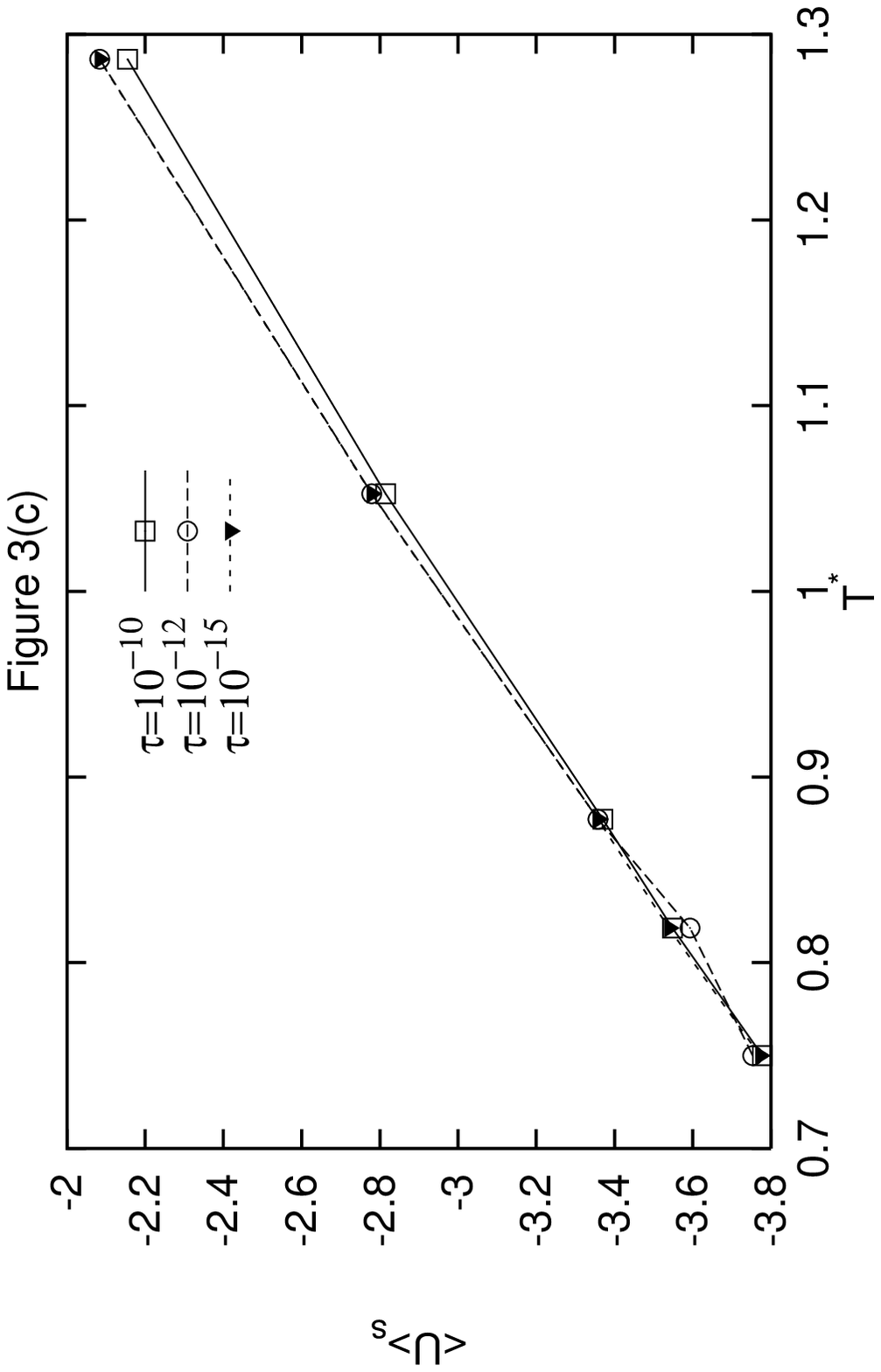}
\end{figure}

\begin{figure}
\includegraphics{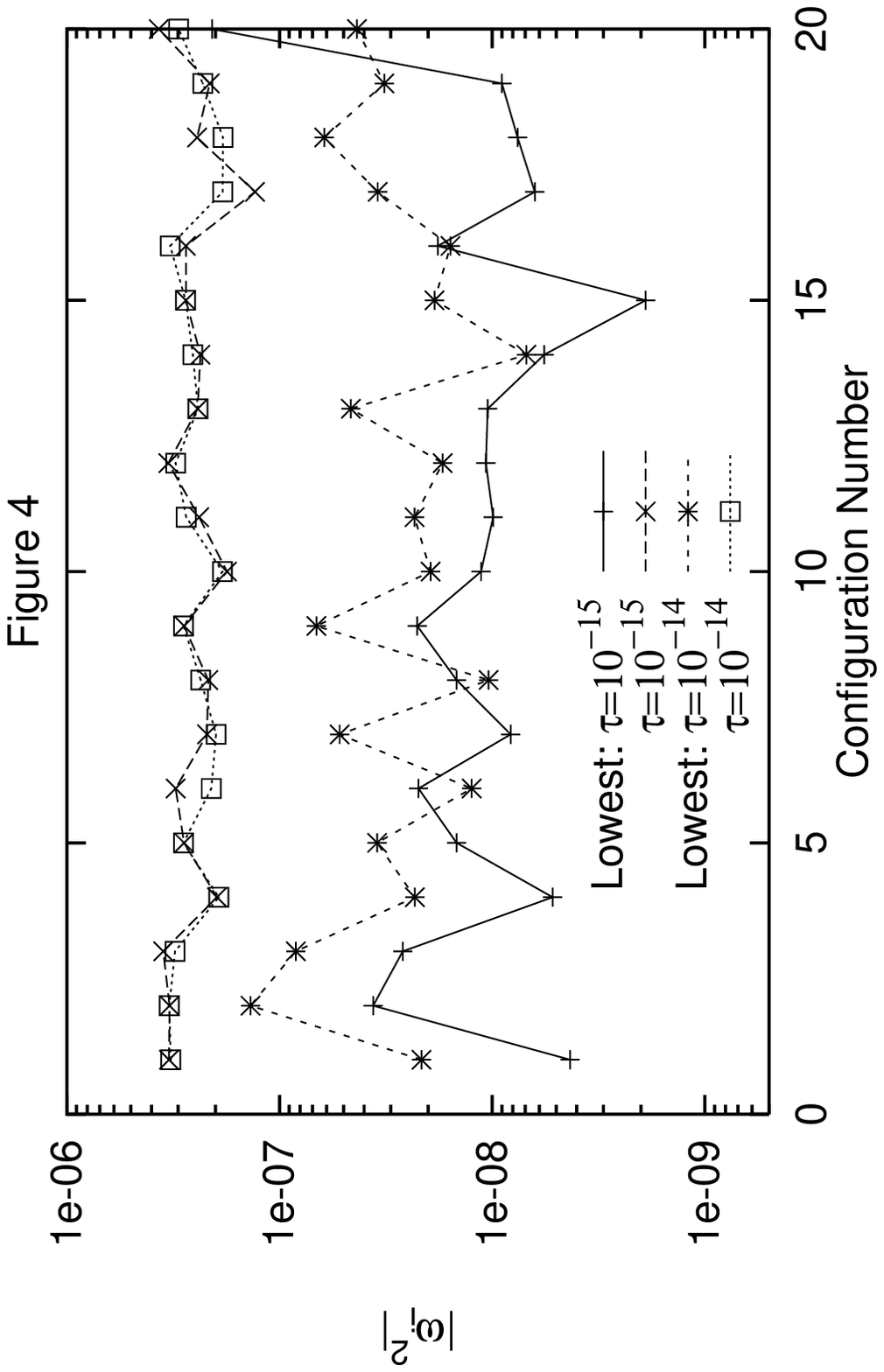}
\end{figure}

\begin{figure}
\includegraphics{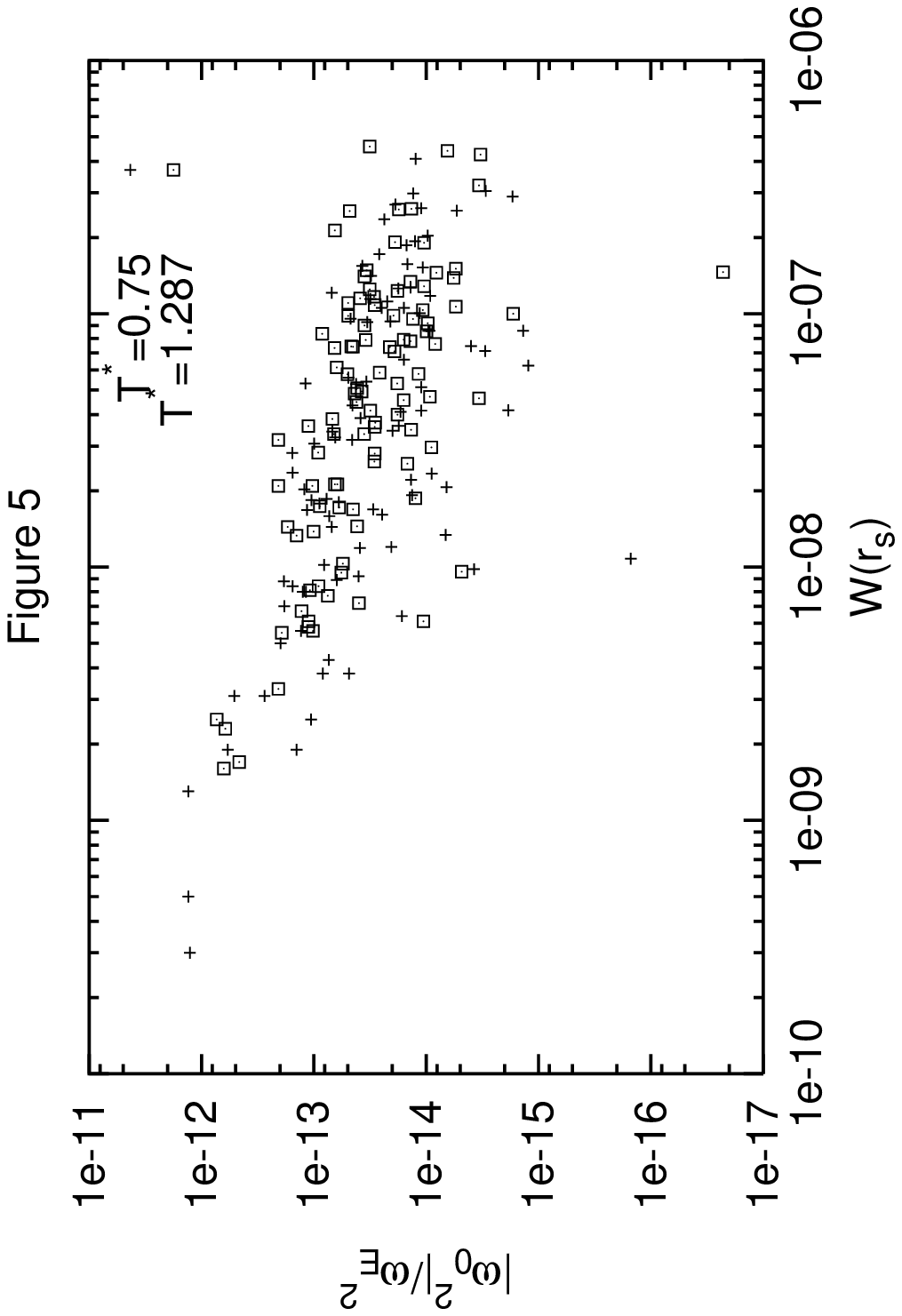}
\end{figure}

\end{document}